\RequirePackage{fix-cm}
\documentclass[twocolumn,final,epjc3]{svjour3}  
\usepackage{cite}
\RequirePackage{graphicx}
\usepackage{isotope}
\usepackage{hyperref}
\usepackage[separate-uncertainty=true]{siunitx} 
\usepackage{color}
\usepackage{amsmath}
\renewcommand \thesection {\Roman{section}}
\RequirePackage{lineno}\newdimen\linenumbersep\linenumbersep=1.5pt\renewcommand{\linenumberfont}\tiny

\newcommand{\dpump}{$\Delta P_{pump}$}
\newcommand{\dpiston}{$\Delta P_{piston}$}
\newcommand{\munster}{Institut f\"ur Kernphysik, Westf\"alische Wilhelms-Universit\"at M\"unster, M\"unster, Germany}
\newcommand{\rpi}{Department of Physics, Applied Physics and Astronomy, Rensselaer Polytechnic Institute, Troy NY, USA}
\newcommand{\stanford}{Physics Department, Stanford University, Stanford CA, USA}

\newcommand{\onebridge}{OneBridge Solutions, Boise, ID, USA}
\newcommand{\wisconsin}{University of Wisconsin - Madison, Madison WI, USA}
\newcommand{\ptbbraunschweig}{Physikalisch-Technische Bundesanstalt, Braunschweig, Germany}

\journalname{Eur. Phys. J. C}
\begin{document}
\title{Magnetically-coupled piston pump for high-purity gas applications}
\author{E.~Brown\thanksref{munster,rpi,brown}
\and
 		A.~Buss\thanksref{munster}
\and
        A.~Fieguth\thanksref{munster}
\and
		C.~Huhmann\thanksref{munster}
\and
		M.~Murra\thanksref{munster} 
\and
		H.-W.~Ortjohann\thanksref{munster}
\and
		S.~Rosendahl\thanksref{munster,ptbbraunschweig}
\and
		A.~Schubert\thanksref{stanford,onebridge}
\and
		D.~Schulte\thanksref{munster, schulte}
\and
		D.~Tosi\thanksref{stanford,wisconsin}
\and
        G.~Gratta\thanksref{stanford}
\and
		C.~Weinheimer\thanksref{munster}
		}	   
\thankstext{brown}{e-mail: browne7@rpi.edu}
\thankstext{schulte}{e-mail:dennyschulte@uni-muenster.de	}
\thankstext{ptbbraunschweig}{Now at \ptbbraunschweig}
\thankstext{onebridge}{Now at \onebridge}
\thankstext{wisconsin}{Now at \wisconsin}
\institute{\munster \label{munster}
		   \and
		   \rpi \label{rpi}
		   \and
		   \stanford \label{stanford}
           }
\date{Received: date / Accepted: date}
%
%
\maketitle
%
\begin{abstract}
Experiments based on noble elements such as gaseous or liquid argon or xenon utilize the ionization and scintillation properties of the target materials to detect radiation-induced recoils. A requirement for high light and charge yields is to reduce electronegative impurities well below the ppb\footnote[1]{Parts per billion, 1\,ppb\,$=1\cdot 10^{-9}$\,mol/mol} level. To achieve this, the target material is continuously circulated in the gas phase through a purifier and returned to the detector. Additionally, the low backgrounds necessary dictate low-Rn-emanation rates from all components that contact the gas.

\begin{sloppypar}Since commercial pumps often introduce electronegative impurities from lubricants on internal components or through small air leaks, and are not designed to meet the radiopurity requirements, custom-built pumps are an advantageous alternative. A new pump has been developed in Muenster in cooperation with the nEXO group at Stanford University and the nEXO/XENON group at Rensselaer Polytechnic Institute based on a magnetically-coupled piston in a hermetically sealed low-Rn-emanating vessel. This pump delivers high performance for noble gases, reaching more than 210 standard liters per minute (slpm) with argon and more than \SI{170}{slpm} with xenon while maintaining a compression of up to \SI{1.9}{bar}, demonstrating its capability for noble gas detectors and other applications requiring high standards of gas purity.\end{sloppypar}
\end{abstract}
\section{Introduction}\label{sec:Introduction}
\begin{sloppypar}Detectors based on noble elements have become widespread in many applications such as Compton telescopes\,\cite{bib:LXeDet,bib:LXeGRIT,bib:LArGO}, ionization calorimeters\,\cite{bib:ElecID,bib:SpaceCharge,bib:MEG}, neutrinoless double-beta-decay searches\,\cite{bib:LArIonization,bib:EXO200Results,bib:nEXOSensitivity,bib:NEXT}, and direct dark matter detection experiments\,\cite{bib:XENON1TResults,bib:LZTDR,bib:DEAP3600,bib:PANDAX,bib:DarkSide20k}. Radiation-induced recoils in the detector medium produce scintillation and ionization signals that are read out by photosensors or charge sensors. As these detectors become larger, the propagation of the light and charge must improve to reach the required threshold and energy resolution, necessitating an increased demand on gas purifying systems. Similarly, backgrounds from radioactive impurities like Rn must be minimized to reach high sensitivity to rare events \cite{bib:nEXOSensitivity,bib:XENON1T}. \end{sloppypar}

\begin{sloppypar}The operational specifications are dominated by the need to drift electrons over lengths 1\,m and greater \cite{bib:nEXOSensitivity,bib:XENON1T}. Electronegative elements like O$_2$ and H$_2$O are continually introduced to the detector material by outgassing of detector components. As these, and other electronegative impurities, impede charge and light propagation, they are continuously removed, usually by pumping the detector material in the gaseous phase through a heated metal getter, then returning it to the detector. In the case of XENON1T, an O$_2$ equivalent concentration in the xenon below 1\,ppb is required to drift charge over a 1\,m scale without appreciable electron attenuation via attachment to impurities \cite{bib:XENON1T}.\end{sloppypar}

The other aspect of detector purity involves Rn mitigation, which is predominantly handled by careful selection of materials with low Rn-emanation rates \cite{bib:XE1TScreening}. The gas handling and purification systems, including the pumps, are a key contributor to the internal Rn background. Incorporating radiopurity screening in a collaborative effort with pump manufacturers provides some level of success, reaching Rn emanation rates of a few mBq, but further Rn reduction by an order of magnitude is still needed. 

At the same time, since detectors become larger, the requirements on pumps increase in kind. Larger detectors need a higher purity to reach the same level of charge attenuation. This is coupled with the fact that there is more material to clean, which necessitates a pump with significantly improved performance to allow high throughput at a pressure differential greater than 1\,bar.

To address these issues, a custom pump was designed for the EXO-200 experiment \cite{bib:EXO200Pump}, which uses a hermetically-sealed magnetically-coupled piston with only high-purity components in contact with the gas. This utilizes a permanent magnet laser-welded into a stainless steel canister to serve as a piston, which is driven using a magnetic ring coupled to a linear drive outside the pump volume. With a cylindrical chamber with a length of \SI{400}{mm} and an inner diameter of \SI{65}{mm} this pump has achieved flows of up to \SI{16}{slpm} at a compression of \SI{1}{bar} for xenon gas. For current ton-scale experiments though, this performance is not adequate.

By scaling the EXO-200 pump to support flow rates above 100\,slpm and pressure differentials greater than 1\,bar, a new, high performance pump has been developed as R\&D for the XENON dark matter project and the nEXO neutrinoless double-beta-decay experiment. This new pump features a larger effective volume with a length of \SI{520}{mm} and an inner diameter of \SI{127}{mm} in combination with an enhanced magnetic gradient based on alternating polarity to allow $\mathcal{O}$(kN) coupling forces between internal and exterior magnets. As a result, our new set-up reached flow rates at $\mathcal{O}$(100)\,slpm at a pressure differential up to 2\,bar.

The design of the magnetic coupling is described in section \ref{sec:MagDesign}, and the mechanical design of this pump is then described in detail in section \ref{sec:MechDesign}. The performance for argon and xenon is then presented in \ref{III}, and section \ref{IV} gives a conclusion and an outlook.

\section{Magnet Design}\label{sec:MagDesign}
To drive a piston using coupling of permanent magnets, an optimized magnet configuration was developed. Since the pressure specifications on this pump are around a factor 2 higher than those of the EXO-200 pump, and since the flow increase of an order of magnitude requires a larger cross sectional area of the pump chamber, a significant improvement in magnetic coupling is needed. 

To this end, magnet configurations were designed using finite element simulations in Comsol \cite{bib:COMSOL} to optimize the magnetic coupling strength between the drive magnets and the piston. The magnetic field \textbf{B} is first found for a given configuration of piston magnets, and the restoring force on the piston \textbf{F} can then be calculated as
\begin{equation}
\textbf{F} = \iiint_V \nabla (\textbf{M} \cdot \textbf{B}) dV,
\label{eqn:Force}
\end{equation}
where \textbf{M} is the magnetization of the magnets, and the integral runs over the volume of all external magnets. 

\begin{figure}[!h]
	\centering\includegraphics[width=.8\linewidth]{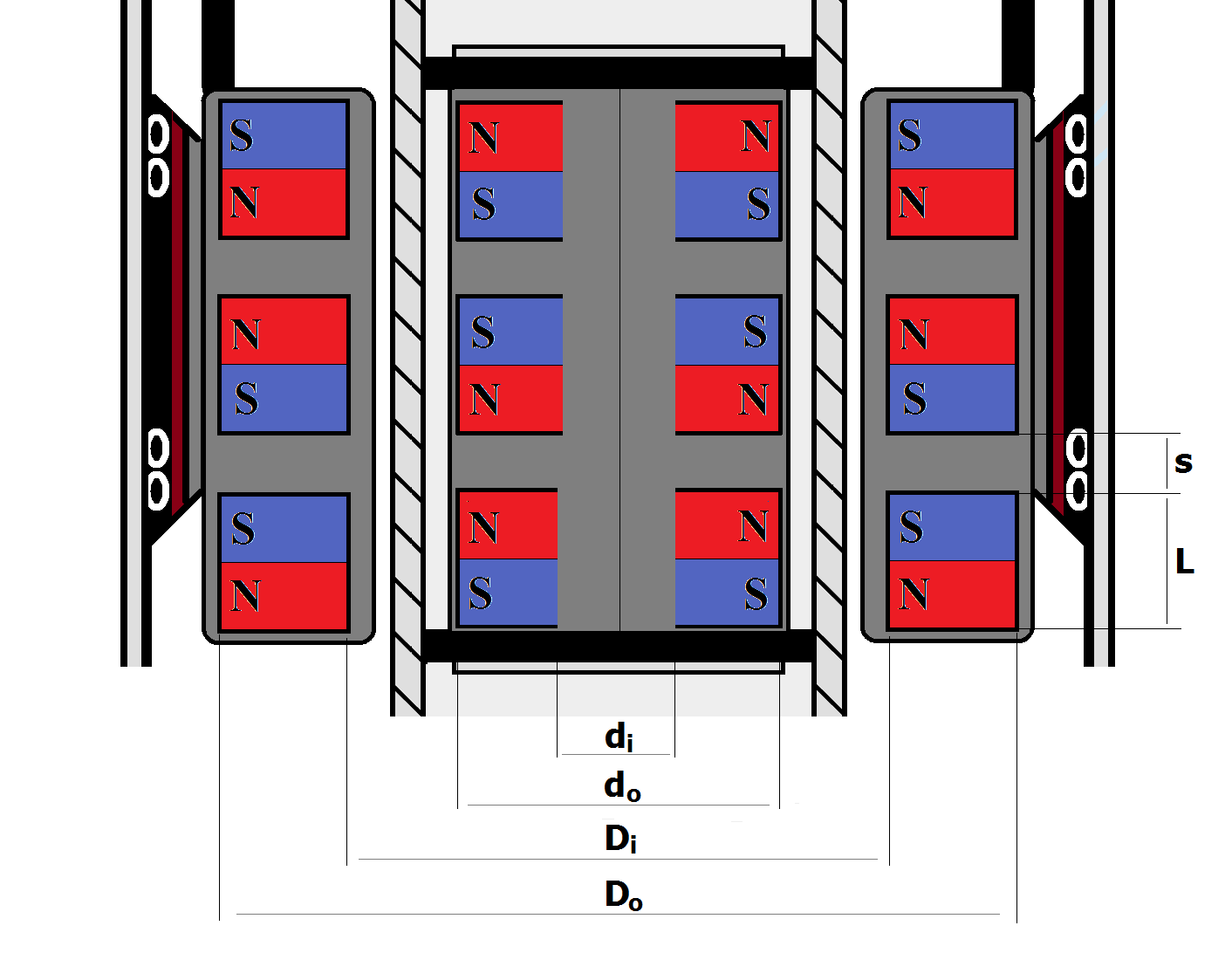}
 	\caption{Cross section of the magnet arrangements considered for the design. The center magnets represent ring magnets inside the piston, and the left and right columns represent the external ring magnets used to drive the piston. The piston magnets alternate with magnetization pointed N-S, S-N, N-S, etc., with the magnetization along the axis of the pump, and with adjacent magnets aligned with like poles together. The drive magnets have the opposite magnetization alignment as the paired piston magnet to form a closed flux loop, and also have alternating magnetization direction along the length of the pump. The inner and outer diameter of the cylindrical inner magnets are denoted by $d_{i}$ and $d_{o}$, while the inner and outer diameter of the external ring is given by $D_{i}$ and $D_{o}$. The length of the magnets is indicated by $L$ and the distance between the ring pairs is indicated by $s$.}
 	\label{fig:magnetschematic}
\end{figure}

The magnet arrangement considered here, as shown in figure \ref{fig:magnetschematic}, uses an alternating orientation of longitudinal magnetization in a row of magnets along a cylinder. The piston magnets are on the cylinder axis, and consist of multiple ring magnets with magnetization along the axis that alternate N-S, S-N, etc., such that the same poles are pointed together for adjacent magnets. The drive magnets are arranged with opposite magnetization direction to that of the piston magnets to form a closed flux loop around each piston magnet. The drive magnets also have like poles pointed together for adjacent magnets. When the piston magnets are each centered within their corresponding drive magnet, the piston is in equilibrium. Upon displacement, there is a restoring force between concentric pairs, and there is an additional cross-coupling from the adjacent drive magnets, which increases the coupling strength non-linearly with the number of magnet pairs. The cross coupling is the key to obtaining the specs needed for the new pump.

\begin{figure}[!h]
	\centering\includegraphics[width=\linewidth]{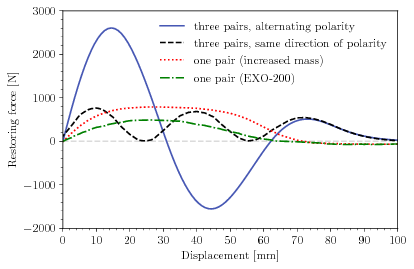}
 	\caption{Simulated restoring force as a function of piston displacement for different magnet configurations. The EXO-200 pump configuration is shown in dash-dotted green. One magnet pair with increased size compared to the EXO-200 design is visualized in dotted red. Dividing the single magnet pair into three pairs with same direction of polarity results in the  dashed black curve. By applying our new method with alternating polarity, the blue curve is obtained. The new configuration yields a maximum coupling strength of \SI{2600}{N}, which is a factor \num{3.3} larger than the initial EXO-200 design.}
 	\label{fig:exo200vsnewconfig}
\end{figure}

\begin{sloppypar}The restoring forces for different magnet configurations are illustrated in figure\,\ref{fig:exo200vsnewconfig}. The configuration for the EXO-200 pump is shown in dash-dotted green. Here, the piston magnet is a single solid cylinder ($d_{i}$\,=\,\SI{0}{mm}, $d_{o}$\,=\,\SI{25.4}{mm}, $B$\,=\,\SI{1.48}{T}) while the external magnet featured diameters of $D_{i}$\,=\,\SI{45}{mm} and $D_{o}$\,=\,\SI{76}{mm} with a field strength of \SI{1.32}{T}. The magnet lengths were $L$\,=\,\SI{51}{mm}. A maximum coupling strength of \SI{490}{N} was achieved. The data points obtained from our simulations match those presented in \cite{bib:EXO200Pump}. 

Figure \ref{fig:exo200vsnewconfig} also includes simulated restoring forces for configurations with an increased magnet mass. Here the piston was changed to a ring magnet ($d_{i}$\,=\,\SI{80}{mm}, $d_{o}$\,=\,\SI{120}{mm}, $B$\,=\,\SI{1.32}{T}) to match realistic designs based on availability commercial products. The external ring magnets measured $D_{i}$\,=\,\SI{137}{mm} and $D_{o}$\,=\,\SI{147}{mm} with a field strength of \SI{1.38}{T}. The red curve shows a single magnet pair with a length of $L$\,=\,\SI{60}{mm}. In spite of the weaker field strength of the inner magnet, there is a stronger coupling force of \SI{790}{N}, due to the increased magnet mass.
The final curves show the same magnet dimensions as for the red curve, but with the length divided into three equal segments measuring \SI{20}{mm}. By spacing the three segments a distance  $s$\,=\,\SI{10}{mm} apart, and orienting the magnets with the same polarity of magnetization, the dashed black curve is obtained. The gain in coupling strength is negligible, but three equilibrium points can be observed. This is the consequence of having three magnet pairs that allow a displacement of one ring magnet in the piston with respect to each of the three outer magnet rings. Using the same magnet placement, but utilizing our new magnet configuration, yields the blue curve. 
The new configuration gives an additional boost of the coupling strength up to \SI{2600}{N}, a factor \num{3.3} higher that the maximum coupling of the EXO-200 pump.\end{sloppypar}

In order to find the optimal case for the new design, a variety of magnet configurations were simulated using our new, alternating technique, and the coupling force was compared.
Based on the availability of commercial magnets, several parameters were fixed and used as constraints in the simulations. The piston magnets considered had a field strength of \SI{1.32}{T} and were solid cylinders with a diameter of $d_{o}$\,=\,\SI{120}{mm}.
The outer ring magnets had the same field strength of \SI{1.32}{T} and had inner and outer diameters that ranged from \SI{130}{mm} to \SI{200}{mm}. The length of the magnet pairs $L$ and the spacing $s$ along the pump axis between adjacent magnet pairs were varied, as was the number of magnet pairs used.

\begin{figure}[!ht]
	\centering
    \vspace{-12pt}
	\includegraphics[width=\linewidth]{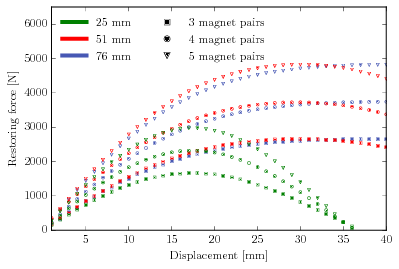}
    \caption{Coupling force as a function of piston displacement for different magnet lengths and numbers of magnet pairs. }
    \label{fig:MagForces}
\end{figure}

The variation of the number and length of the magnets was performed simultaneously in order to directly compare the difference between the same magnet mass in different arrangements. For example, 2 magnet pairs of a given length have the same mass as 4 magnet pairs that are half as long. In these simulations, magnet lengths $L$ of 
\SI{25}{mm}, \SI{51}{mm}, and \SI{76}{mm} were considered, with the number of magnet pairs ranging from 3 to 5 and an inter-magnet spacing of $s$\,=\,\SI{10}{mm}. The drive magnets used in these simulations had an inner diameter of $D_{i}$\,=\,\SI{140}{mm} and an outer diameter of $D_{o}$\,=\,\SI{200}{mm}. The resulting coupling force for a subset of these simulations is shown as a function of piston displacement in figure \ref{fig:MagForces}. 

Several trends are apparent from this study. First, there is little distinction between the \SI{51}{mm} and \SI{76}{mm} magnets for the same number of magnet pairs, indicating that the longer magnets become ineffective at increasing the coupling. Additionally, there is a clear increase in coupling strength with increasing numbers of magnets, as expected naively. 
The most important feature shown in this figure is the increase in coupling force due to the cross coupling of adjacent magnet pairs. This is evident when comparing the peak restoring force for different magnet masses. For example, the curve for five \SI{25}{mm} magnets peaks at a higher force than the curve for three \SI{51}{mm} magnets, in spite of the fact that it has less magnet mass. Finally, these results demonstrate the feasibility of reaching $\mathcal{O}$(kN) forces with a modest number of magnets.

\begin{figure}[!ht]
	\centering
    \includegraphics[width=\linewidth]{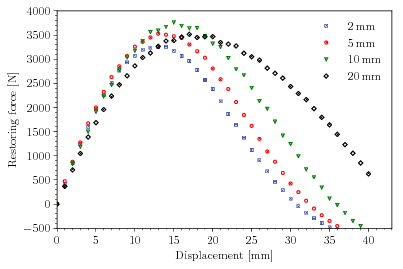}
    \caption{Magnetic coupling vs displacement for arrangements with different spacing distances between magnet pairs.}
    \label{fig:Spacing}
\end{figure}

The next optimization considered was the spacing $s$ along the pump axis between adjacent magnet pairs. In these simulations, two magnet pairs were used with piston magnets measuring $L$\,=\,\SI{20}{mm} in length and drive magnets with an inner diameter of $D_{i}$\,=\,\SI{140}{mm} and an outer diameter of $D_{o}$\,=\,\SI{200}{mm}. The spacing distance $s$ was then varied from \SI{2}{mm} to \SI{20}{mm}. As shown in figure \ref{fig:Spacing}, there is a peak in the restoring force at a \SI{10}{mm} spacing indicating that this is the ideal spacing size. The same optimum at \SI{10}{mm} spacing was seen for other magnet lengths, suggesting that this optimum point is related to the fixed diameter of the piston magnets. 

\begin{figure}[!ht]
	\centering
    \includegraphics[width=\linewidth]{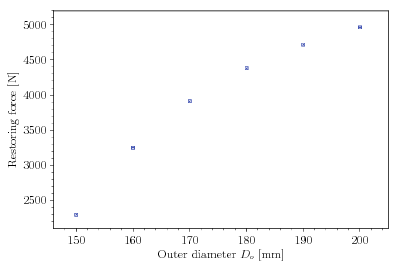}
    \caption{Maximum coupling for drive magnets with an inner diameter of \SI{140}{mm} and varying outer diameters.}
    \label{fig:OuterD}
\end{figure}

Studies were then performed on the size of the drive magnets to find the optimal inner and outer diameters. For the outer diameter studies, the inner diameter was fixed at $D_{i}$\,=\,\SI{140}{mm}, and the outer diameter $D_{o}$ was varied from \SI{150}{mm} to \SI{200}{mm}. As shown in figure \ref{fig:OuterD}, there is a modest increase by about a factor of 2 at the extremes. But this should be compared to the change in magnet mass, which increases by a factor of 7. There is also a slight flattening in the slope when $D_{o}$ is around \SI{180}{mm}, indicating magnets larger than this only add minimally to the coupling strength.

\begin{figure}[!ht]
	\centering
    \includegraphics[width=\linewidth]{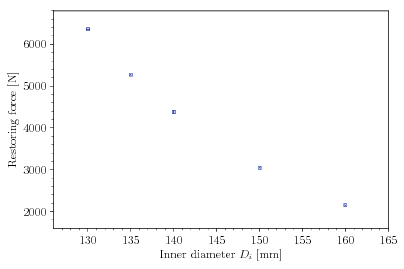}
    \caption{Maximum coupling for different inner diameters of the drive magnets.}
    \label{fig:InnerD}
\end{figure}

\begin{sloppypar}To study the impact of the inner diameter $D_{i}$, the thickness of the outer ring was fixed at \SI{20}{mm}. All other parameters were fixed to the same values as for the outer diameter studies, and the inner diameter $D_{i}$ was varied from \SI{130}{mm} to \SI{160}{mm}. Figure \ref{fig:InnerD} shows the maximum coupling, which has a steeper slope than that of  the outer diameter $D_{o}$ . This is because the magnetic field varies most strongly close to the piston magnets, so to achieve the strongest coupling the outer magnets should be placed as close as possible to the piston.\end{sloppypar} 

\begin{figure}[h!]
\centering
\includegraphics[width=8cm]{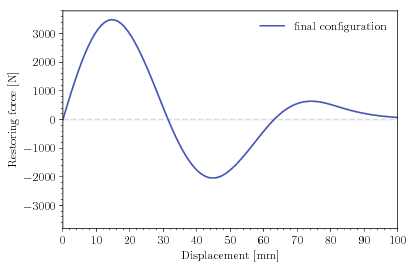}
\caption{Simulation of the restoring force versus ring displacement of the final pump configuration. A maximum of \SI{3500}{N} can be seen.}
\label{fig:Coupling}
\end{figure}

The features seen in these studies were employed in the development of the pump presented here. By using multiple, short magnets with optimized spacing and geometry, a significant increase in coupling strength is possible relative to the EXO-200 pump upon which this is based. Simulations were conducted on the configuration chosen for the pump developed for this work. The piston was chosen with three ring magnets measuring $d_{i}$\,=\,\SI{80}{mm}, $d_{o}$ = \SI{120}{mm}, and a $L$\,=\,\SI{20}{mm}, and with a field strength $B\,=\,\SI{1.32}{T}$. These are arranged with our alternating polarity design with a distance between each ring of $s$\,=\,\SI{10}{mm}. The outer portion contains three matching magnet rings with the same linear dimensions and diameters of $D_{i}$\,=\,\SI{137}{mm}, $D_{o}$\,=\,\SI{157}{mm}, and a field strength of $B\,=\,\SI{1.38}{T}$. Figure \ref{fig:Coupling} shows the predicted coupling force for this configuration as a function of displacement of the piston relative to the outer magnet rings. 
This configuration yields a maximum coupling force of \SI{3500}{N}, a factor of \num{7} larger than the initial EXO-200 design as presented in figure\,\ref{fig:exo200vsnewconfig}.
As described in section \ref{sec:MechDesign}, this coupling strength is sufficient to provide a pressure differential of up to 2 bar with a large aperture pump volume.

\section{Mechanical Design}\label{sec:MechDesign}
\begin{figure}[h!]
\centering
\includegraphics[width=8cm]{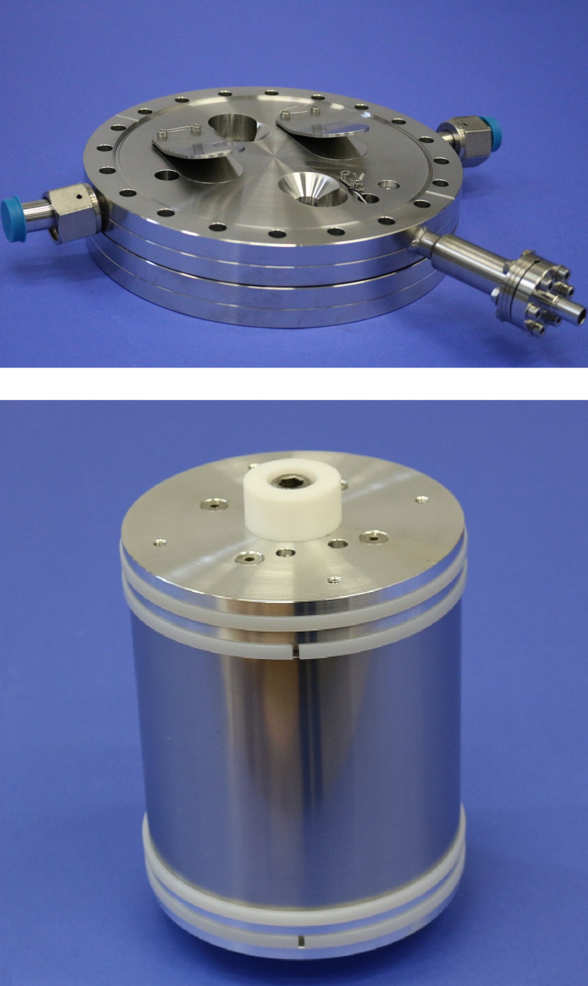}
\caption{Top: End flange containing gas-inlet and outlet, temperature sensor, feed-through and flapper valves. Bottom: Magnetically-coupled, laser-welded piston with gaskets.}
\label{fig:flangepiston}
\end{figure}

The magnetically-coupled piston pump was constructed of a monolithic type 316L stainless-steel (SS) body with a length of \SI{520}{mm}, an inner diameter of \SI{127}{mm}, and an outer diameter of \SI{133}{mm}, giving a total volume of \SI{4.5}{l}. The inner wall of the cylinder was honed to create a low-friction surface. Each end is closed with custom ConFlat SS flanges to allow gas to enter and exit the volume and to provide a vacuum port for cleaning (top in figure \ref{fig:flangepiston}). These are sealed with copper gaskets to maintain high leak-tightness. The gas ports consist of electropolished SS tubes with VCR connections that use metal gasket seals. These are welded directly to the custom ConFlat flanges on either end of the pump. There are three ports on each end of the pump to serve as an inlet and outlet by use of flapper valves, and an open line to serve as an unimpeded port for pumping vacuum on the pump to clean it before use and to measure the pressure inside the pump body.

\begin{sloppypar}Flapper valves constructed from sub-millimeter spring-steel foils maintain unidirectional flow. These are supported with an SS grid on one side to prevent the flappers from popping into the vent port and with an SS plate on the other side to prevent excess bending that could damage the flappers. The flapper valves allow both ends of the pump to alternatively supply compression at the gas discharge and expansion to draw in low pressure gas. Additionally, the top flange contains a PT1000 temperature sensor wired with vacuum compatible PTFE coated cable and connected to a ceramic CF16 feedthrough.\end{sloppypar}

The piston consists of a set of three rings of permanent neodymium magnets with a strength of \SI{1.32}{T}, each with a length $L$ of \SI{20}{mm}, an inner diameter $d_{i}$ of \SI{80}{mm}, and an outer diameter $d_{o}$ of \SI{120}{mm}. The magnetization of the rings is oriented along the axis of the pump, but with alternating direction, as described in section \ref{sec:MagDesign}. The magnets are supported by a custom non-magnetic aluminum support structure that maintains a \SI{10}{mm} gap between each ring magnet. This entire assembly is contained in an SS vessel with a length of \SI{155}{mm} and outer diameter of \SI{125}{mm} (bottom in figure \ref{fig:flangepiston}), which is hermetically sealed via laser welding to ensure no contact between the gas and the magnets. The piston is sealed against the inner wall of the pump volume with ultra high molecular weight polyethylene (UHMWPE) gaskets that are connected with SS mounts. Thus, all internal components that come into contact with the gas consist solely of vacuum compatible materials, which minimizes contamination via outgassing and Rn emanation.
While surface treatment, such as electropolishing, is known to further reduce Rn emanation from materials, this was not done for internal components of the pump, as the focus was to create a working prototype with the elimination of the high Rn emanating material used in commercial products. Such surface treatment is one potential area for improvement in derivative devices.

The gasket design has been changed compared to \cite{bib:EXO200Pump} to create a larger sealing contact between piston and cylinder wall. The new gasket design is also optimized to minimize the dead volume between the endcap of the pump body and the end of the piston, which allows a more complete exhaust of gas in each stroke. This improves performance and reduces adiabatic heating of gas that remains inside the pump body.

Due to the fact that the gaskets are directly in contact with the pure noble gas, the material has to have low rates of outgassing, Rn emanation, and physical wear. Tests performed in \cite{bib:EXO200Pump} showed that UHMWPE performed adequately in these regards, and as such was used for the gaskets in this pump (bottom in figure \ref{fig:flangepiston}).
By utilizing a vertical orientation of the pump body, this leads to a more symmetrical alignment of the piston and further reduces the wear rate.

\begin{figure}[h!]
\centering
\includegraphics[width=8cm]{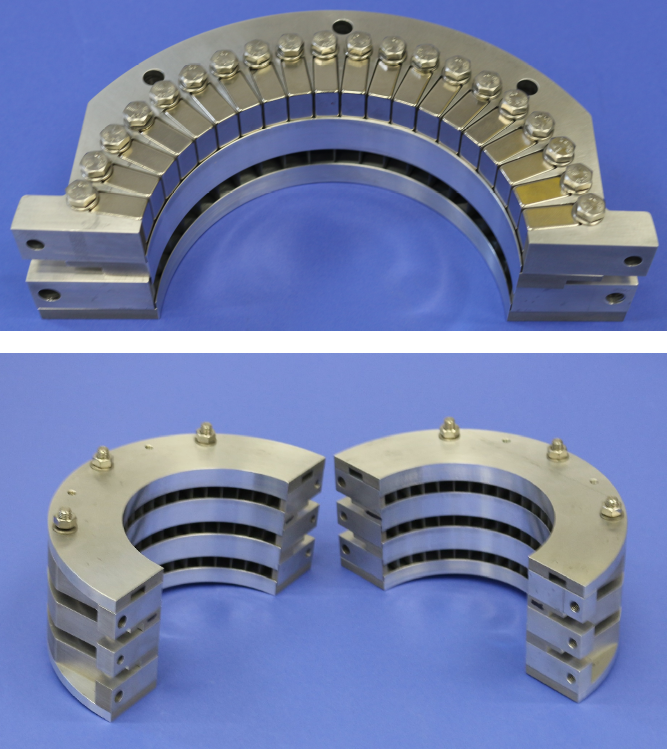}
\caption{Top: Sequence of bar magnets for outer magnetic ring. Bottom: Two halves of the complete outer magnetic ring.}
\label{fig:magnets}
\end{figure}

The piston is coupled magnetically to a set of magnetic rings located outside the pump volume. These are constructed of permanent neodymium bar magnets with a strength of \SI{1.38}{T}  
measuring \SI{20}{mm} $\times$ \SI{10}{mm} $\times$ \SI{20}{mm} in a cylindrical arrangement, with the \SI{10}{mm} dimension tangent to the circumference These are supported by a custom nonmagnetic aluminum frame as shown in figure \ref{fig:magnets}. The frame holds the three outer magnet rings in equal and opposite magnetization orientation to the matching three magnet rings in the piston, again along the axis of the pump, and with the same \SI{10}{mm} spacing between the individual rings. The inner diameter $D_{i}$ of the outer magnet frame is \SI{137}{mm}, yielding an \SI{8.5}{mm} radial gap between the inner and outer magnets.

This configuration yields a maximum coupling force of \SI{3500}{N}, and is in excellent agreement with measurements of the coupling, which yielded a maximum coupling force of \SI{3468(28)}{N}. This coupling strength corresponds to a pressure difference of about \SI{2.7}{bar} across the piston.

The entire assembly is mounted vertically, and the outer magnet assembly is driven with a linear drive, composed of an electric cylinder (SEW, CMS) that is powered by a frequency converter (SEW, MDX 61B). The converter uses a Modbus connection, allowing operation via an external slow control environment. In our case this was designed in LabVIEW.

\begin{figure}[h!]
\centering
\includegraphics[width=8cm]{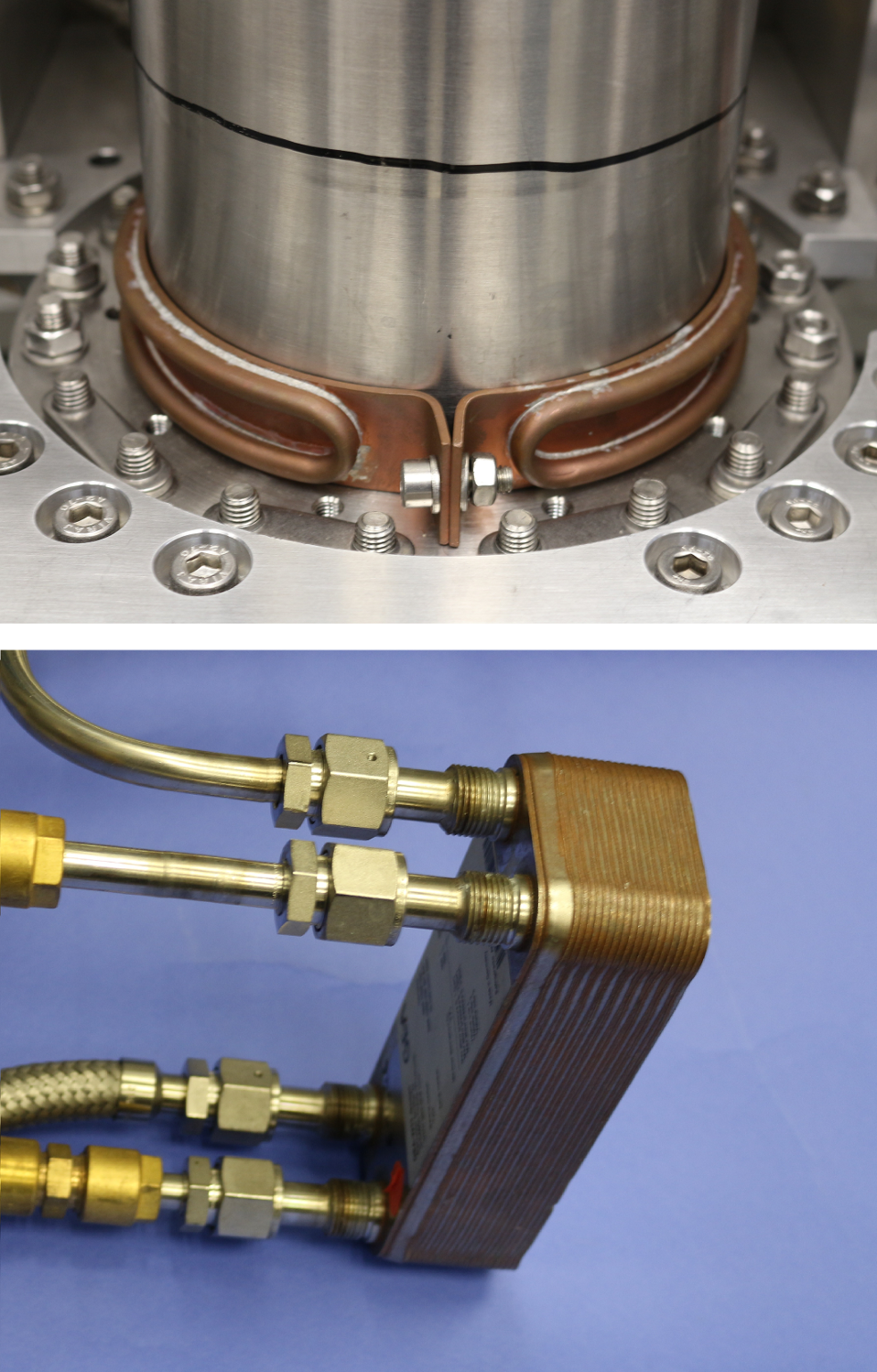}
\caption{Top: Water cooling system for the pump body. Copper shells flushed with cold water cool the space of highest compression directly beneath top flange and above bottom flange. Bottom: A heat exchanger pre-cools the gas before it enters the pump.}
\label{fig:cooling}
\end{figure}

In order to avoid demagnetization of the neodymium magnets at critical temperatures higher than \SI{70}{\celsius} during continuous operation\,\cite{bib:HCKM}, a cooling system was integrated in the system as shown in figure\,\ref{fig:cooling}. Copper shells are attached to the endcaps of the body, which are water-cooled to provide cooling in the compression volume. Additionally, a counterflow heat exchanger is used to pre-cool the gas with the same cooling water before it enters the pump. 

\begin{figure}[!ht]
\centering
\includegraphics[width=.8\linewidth]{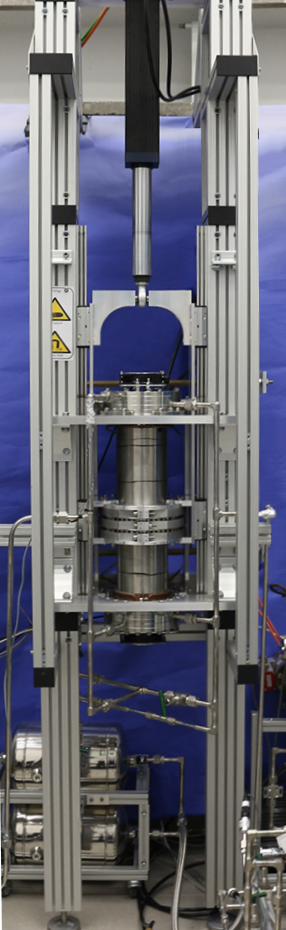}
\caption{Fully assembled pump at Muenster pump test station.}
\label{fig:pump}
\end{figure}

The fully assembled pump at the Muenster pump-testing station can be seen in figure\,\ref{fig:pump}.

\section{Performance and longterm stability}\label{III}
\begin{figure}[h!]
\centering
\includegraphics[width=8.7cm]{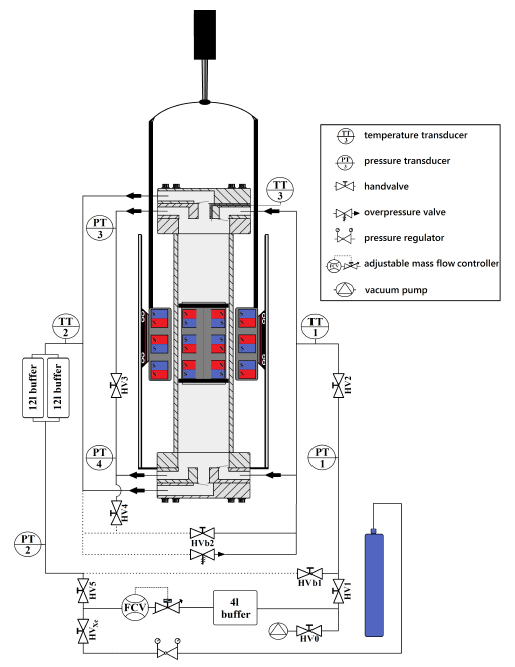}
\caption{Flow chart of the pump at Muenster pump test station. The black arrows indicate the unidirectional flow through the flapper valves. The red and blue color of the different magnets indicate their axial polarities.}
\label{fig:teststation}
\end{figure}

The performance of the pump was tested at the Muenster pump-test station, which is shown schematically in figure \ref{fig:teststation}. The entire system is evacuated to high vacuum with a scroll pump and turbomolecular pump to the level of 10$^{-7}$~mbar. Xenon or argon is then introduced to the system by a gas bottle with a pressure regulator to allow different pressures in the system, and the gas can be recovered to the bottle via cryopumping by cooling the bottle with liquid nitrogen. The testing circuit includes a mass flow controller (FCV, MKS, 1579A) to measure gas flow up to \SI{211}{slpm} for argon and \SI{200}{slpm} for xenon. Two buffer volumes of \SI{12}{l} and one of \SI{4}{l} are added to reduce flow oscillations due to the small volume of the gas system relative to that of the pump. Additionally, three temperature transducers (TT, Farnell, HEL-705) and four pressure transducers (PT, Swagelok, PTU) are mounted.

Temperatures are monitored by the three PT1000 temperature sensors in combination with an internal temperature sensor of the linear drive. TT1 is installed at the gas inlet, TT2 at the gas outlet and TT3 is inside the pump volume just below the top flange. TT3 is of crucial importance since it is mounted at the point of highest gas compression and thus at the point of highest temperature. 

The pressure sensors PT1 and PT2 measure the pressures before and after the pump. The differential pressure across the pump is then defined as $\Delta P_{pump} = \text{PT}2 - \text{PT}1$ and measures the pressure difference between the inlet and outlet of the pump. PT3 and PT4 are located next to the compression volumes of the pump and can therefore measure pressure differential across the piston as  $\Delta P_{piston} = \text{PT}3 - \text{PT}4$, or the pressure above the top of the piston minus the pressure below the bottom of the piston. With this definition, a positive $\Delta P_{piston}$ corresponds to an upward stroke, while negative values correspond to a downward stroke. HV3 is closed during normal operation to isolate the two ends of the pump, and is only opened for cleaning under vacuum.

During standard operation, if $\Delta P_{piston}$ exceeds the critical decoupling pressure of $\Delta P_{crit}=$\,\SI{2.7}{bar}, when force of the gas against the piston plus the friction of the gaskets against the pump wall is greater than the magnetic coupling force, the piston will decouple from the outer ring. As shown in figure \ref{fig:Coupling}, there are two stable equilibrium points, where the restoring force is zero and the slope is positive. This indicates that upon decoupling, the piston will lag the outer magnets by about 60 mm. This is not inherently a problem, since the piston will restore to alignment at the end of the corresponding return stroke. Nevertheless, this is avoided by operating the pump below decoupling pressure. A safety margin is used to account for the friction, and $\Delta P_{piston}$ is kept below \SI{2.2}{bar}.

\begin{figure}[h!]
\centering
\includegraphics[width=8cm]{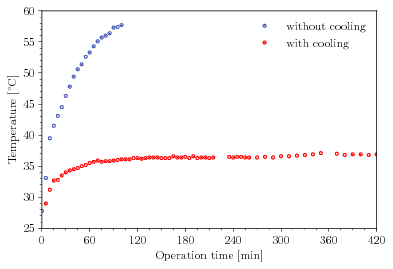}
\caption{Heat evolution inside the pump without cooling system (blue) and with cooling system (red) at operation with argon at an inlet pressure of \SI{1.45(5)}{bar}, a flow of \SI{103(4)}{slpm} and a \dpiston~of \SI{0.76(5)}{bar}. This test was performed with a prototype external magnet ring with lower field strength.}
\label{fig:temperatures}
\end{figure}

First tests were performed to monitor the temperature evolution during operation using argon at an inlet pressure of \SI{1.45(5)}{bar}, a flow of \SI{103(4)}{slpm} and a \dpump~of \SI{0.76(5)}{bar}. Figure \ref{fig:temperatures} shows the heat evolution inside the pump for the argon tests at TT3 with and without the cooling system. Due to the temperature exceeding \SI{50}{\celsius} without the cooling system, the pump could not be operated stably, and the test was stopped to avoid overheating the magnets. However the operation of the cooling system adequately reduced the temperature inside the pump to around \SI{35}{\celsius} for long term operation. This was tested with a prototype external magnet ring with lower field strength, but was still sufficient to demonstrate the functionality of the cooling system. This is verified by the long term stability tests presented below.

\begin{figure}[h!]
\centering
\includegraphics[width=8cm]{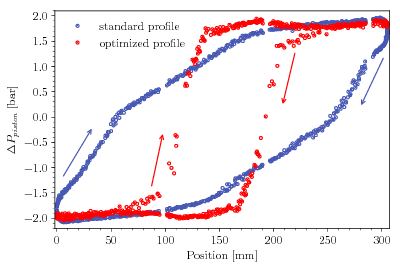}
\caption{Piston pressure \dpiston~vs the position of the external ring using xenon gas with a mean inlet pressure of \SI{1.8(1)}{bar}. The optimized drive profile (red) allows to build up and hold a high piston pressure much faster and longer compared to the linear profile (blue). The arrows indicate the movement of the piston during a full stroke up and down.}
\label{fig:profiles}
\end{figure}

Due to the fact that the pressure rises slowly over the piston stroke, maintaining a constant \dpiston~is non trivial. A linear driving profile (standard profile) which drives the piston at a constant speed over the stroke, is inefficient as the maximum \dpiston~is reached very late in the stroke. This was demonstrated using xenon gas at an inlet pressure of \SI{1.8(1)}{bar} and measuring the maximum flow and pressure differential. As shown in figure \ref{fig:profiles}, the monotonically increasing pressure over each stroke yields a non-uniform pressure (blue). 

An optimized profile was implemented that quickly compresses the gas upon turnaround of the piston by moving it at high speed until the operating pressure is reached. The profile then holds the pressure constant over the remainder of the stroke by moving at a slower, constant speed. This is shown in figure \ref{fig:profiles}, which demonstrates a more uniform pressure differential (red).

\begin{figure}[h!]
\centering
\includegraphics[width=8cm]{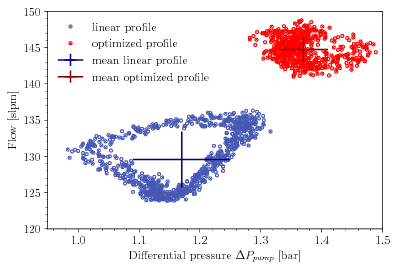}
\caption{Differential pressure \dpump~versus gas flow for the same measurement as that of figure \ref{fig:profiles} with xenon gas at a mean inlet pressure of \SI{1.8(1)}{bar}. The optimization of the drive profile leads to clear improvements of the performance. A higher flow of \SI{144(2)}{slpm} as well as a higher differential pressure of \dpump~=\SI{1.37(4)}{bar} at same maximum piston pressure has been achieved compared to the standard profile, which reached a flow of \SI{129(4)}{slpm} and a differential pressure \dpump~=\SI{1.17(8)}{bar}. Additionally, the oscillations of flow and differential pressure have been decreased as illustrated by the smaller variation of the measurement points.}
\label{fig:flowprofiles}
\end{figure}

\begin{sloppypar}
Further improvement via profile optimization is show in figure \ref{fig:flowprofiles}, which shows the flow vs \dpiston~for the two driving profiles used. The performance improvement is seen both in the absolute flow and \dpump, and also results in a tighter distribution over the pump motion. With the standard profile, a mean flow of \SI{129(4)}{slpm} and a compression of \dpump~=\SI{1.17(8)}{bar} was achieved in contrast to the mean of the optimized profile with a flow of \SI{144(2)}{slpm} and a \dpump~of \SI{1.37(4)}{bar}. As a uniform pressure and flow are usually the most important parameters for stability of liquid noble detectors, this optimization is key to the performance of the pump. \end{sloppypar}

\begin{figure}[h!]
\centering
\includegraphics[width=8cm]{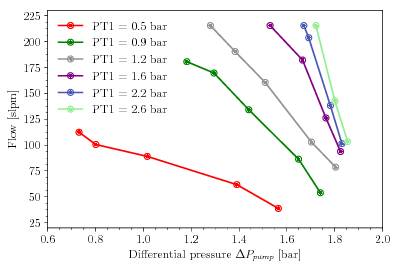}
\caption{Performance of the pump with argon for different inlet pressures. The flow was limited by the maximum point of the mass flow controller for inlet pressures of \SI{1.2}{bar} and higher.}
\label{fig:ArResults}
\end{figure}

\begin{figure}[h!]
\centering
\includegraphics[width=8cm]{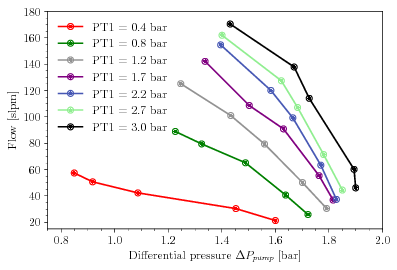}
\caption{Performance of the pump with xenon for different inlet pressures.}
\label{fig:XeResults}
\vspace{-12pt}
\end{figure}

To characterize the performance of the pump quantitatively, measurements of flow vs differential pressure were made using both xenon and argon at different inlet pressures. The inlet pressure can be controlled by closing HV2 within the recirculation circuit stepwise, by varying the amount of the gas in the system, or by changing the linear drive velocity. Thereby, the pump performance can be tested for a wide operation range. Tests were made at a constant pump inlet pressure, as measured by PT1, since this is the simplest pressure to hold constant in our system. For all measurements, the pump was operated at maximum \dpiston, which is the condition of maximal flow. Multiple measurements were then made of the flow and \dpump. The results of the flow and \dpump~are averaged over several strokes of the pump, and are shown for argon in figure \ref{fig:ArResults} and for xenon in figure \ref{fig:XeResults}. The performance was measured for up to seven different inlet pressures at up to five handvalve positions for each inlet pressure. 

For argon, flows exceeded the capacity of the mass flow controller which has a maximum flow for argon of \SI{211}{slpm}. This was achieved for several inlet pressures ranging from \SI{1.2}{bar} to \SI{2.6}{bar}. In particular, at a \SI{2.6}{bar} inlet pressure, a maximum differential pressure of \dpump~= \SI{1.85}{bar} was reached at a flow of \SI{95}{slpm}. At the maximum flow allowed by the mass flow controller of \SI{211}{slpm}, a differential pressure of \SI{1.72}{bar} was obtained for the same inlet pressure.

\begin{sloppypar} For xenon a maximum flow of \SI{171}{slpm} was achieved with a $\Delta P_{pump}$ of \SI{1.45}{bar} at an inlet pressure of \SI{3.0}{bar}. The maximum differential pressure of \dpump~= \SI{1.9}{bar} was reached with a flow of \SI{45}{slpm} at the same inlet pressure of \SI{3.0}{bar}.\end{sloppypar}

The steep flow vs \dpump~curves for argon and xenon at high inlet pressure show that the pump works essentially as a pressure amplifier in this regime, with the flow largely dominated by the impedance of the circuit. This is consistent with the fact that the pumping mechanism is effectively adiabatic compression, with the subsequent motion of the gas being passive flow through the impeding circuit. This feature is more extreme in the argon data. This, coupled with the lower flow rates of xenon compared to argon, is indicative of the increased difficulty of pumping a heavy gas like xenon. Due to this, performance with lighter gases like neon is expected to be even better than for argon.

In contrast, the relatively flat curves for both argon and xenon at low inlet pressure indicate a different performance regime. At inlet pressures below \SI{1.2}{bar} for argon and \SI{0.8}{bar} for xenon, the limited power of the linear drive was insufficient to drive the piston fast enough to reach maximum \dpiston, resulting in a characteristically different flow to pressure relationship.

We observe that performance improves with higher inlet pressure. The anti-correlation between flow and \dpump~should be considered when designing systems to be used with this type of pump. Applications requiring a large pressure differential should be designed with lower flow requirements, and those with high flow requirements should be built with low impedance circuits. 

\begin{figure}[h!]
\vspace{-12pt}
\centering
\includegraphics[width=8cm]{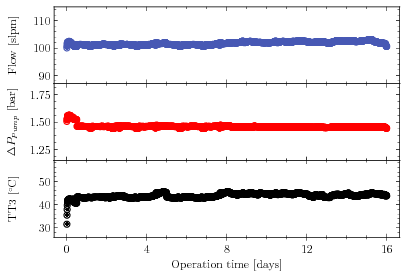}
\caption{Demonstration of longterm stability of the pump using xenon at an inlet pressure of \SI{1.45}{bar}. The performance parameters flow (blue), differential pressure (red) and temperature at the compression volume beneath the top flange (black) are shown.}
\label{fig:stability}
\vspace{-12pt}
\end{figure}

Another important pump characteristic is longterm stability with high performance, as this is required for most applications. To this end, a study of the stability in a high performance state was performed using xenon. Figure \ref{fig:stability} shows the stability of the performance parameters, flow and $\Delta P_{pump}$, as well as of the important temperature TT3 inside the pump just below the top flange.

\begin{sloppypar}A stable operation with an average flow of \SI{100(2)}{slpm} (blue) and a mean differential pressure of \SI{1.42(4)}{bar} (red) has been achieved over a period of \SI{16}{days}. A stable temperature inside the body of \SI{43(1)}{\degreeCelsius} (black) has also been obtained, which is well below the demagnetization temperature for the magnets. Thus, the performance of the cooling system is verified for the final magnet configuration, as well as over a two week long operation.\end{sloppypar}

For some applications, such as rare event experiments, the radon emanation of pumps is of crucial importance. To assess this aspect of the pump, a measurement was performed to determine the emanation rate of $^{222}$Rn from the interior of the completed pump. A gas sample was extracted and measured following the method in \cite{bib:RnEmanation}, resulting in an emanation rate of \SI{330(60)}{\mu Bq}, an order of magnitude cleaner than what has been achieved in commercial pumps. This meets the current radiopurity needs for low background environments. Should further Rn reduction be necessary, surface treatments to minimize the surface area and Rn emanation could be implemented.

\section{Conclusion}\label{IV}
\begin{sloppypar}A new, high-performance pump with a special magnetically-coupled drive mechanism based on an alternating magnet configuration was developed for noble gas applications. The complete isolation of the drive from the gas and the usage of only clean components yields a high purity, with a $^{222}$Rn emanation rate of \SI{330(60)}{\mu Bq}. A stable performance of more than \SI{210}{slpm} for argon and more than \SI{170}{slpm} for xenon combined with a compression up to \SI{1.9}{bar} makes the pump a promising tool for many noble gas and high purity applications. \end{sloppypar}

\begin{acknowledgements}
We would like to express our special thanks to the group of M. Lindner at MPIK Heidelberg, especially to H. Simgen, for determining the radon emanation of our pump. We acknowledge the help of the Electronic Workshop of the Institut f\"ur Kernphysik at M\"unster University, especially of R. Berendes.
Different parts of the pump development at Muenster University have been supported by DFG (WE1843/7-1), Helmholtz Association (HAP) and BMBF (05A14PM1, 05A17PM2).
The work at Stanford was supported, in part, by DoE Offices of High Energy and Nuclear Physics under grants DE-SC0009841 and DE-SC0017970.
The Rensselaer work was supported in part by the NSF Division of Physics under grant No. 1719259.
\end{acknowledgements}
\bibliographystyle{aip2}

\bibliography{references}

\end{document}